# Electron-phonon coupling and superconductivity-induced distortions of the phonon lineshape in V$_3$Si

*A. Sauer[1], D. A. Zocco[1,2], A.H. Said[3], R. Heid[1], A. Böhmer[1], F. Weber[1]*

[1] *Institute for Solid State Physics, Karlsruhe Institute of Technology, 76021 Karlsruhe, Germany*
[2] *Institute of Solid State Physics, Vienna University of Technology, 1040 Vienna, Austria*
[3] *Advanced Photon Source, Argonne National Laboratory, Lemont, Illinois, 60439, USA*

**Phonon measurements in the A15-type superconductors were complicated in the past because of the unavailability of large single crystals for inelastic neutron scattering, *e.g.* in the case of Nb$_3$Sn, or unfavorable neutron scattering properties in the case of V$_3$Si. Hence, only few studies of the lattice dynamical properties with momentum resolved methods were published, in particular below the superconducting transition temperature $T_c$. Here, we overcome these problems by employing inelastic x-ray scattering and report a combined experimental and theoretical investigation of lattice dynamics in V$_3$Si with the focus on the temperature-dependent properties of low-energy acoustic phonon modes in several high-symmetry directions. We paid particular attention to the evolution of the soft phonon mode of the structural phase transition observed in our sample at $T_s = 18.9$ K, *i.e.*, just above the measured superconducting phase transition at $T_c = 16.8$ K. Theoretically, we predict lattice dynamics including electron-phonon coupling based on density-functional-perturbation theory and discuss the relevance of the soft phonon mode with regard to the value of $T_c$. Furthermore, we explain superconductivity-induced anomalies in the lineshape of several acoustic phonon modes using a model proposed by Allen et al. [Physical Review B 56, 5552 (1997)].**

PACS numbers: 71.45.Lr, 63.20.kd, 63.20.dd, 63.20.dk

## I. Introduction

Until the discovery of copper-oxide superconductors in 1986, superconducting compounds featuring the *A15* structure [1] held the record transition temperature with the highest $T_c = 23$ K for Nb$_3$Ge [2,3] and even nowadays Nb$_3$Sn is the most commonly used material for manufacturing high-field magnets (see a recent review by Stewart [1]). In the compounds Nb$_3$Sn and V$_3$Si with $T_c \approx$ 18 K and 17 K, respectively, there is clear evidence of a martensitic cubic-to-tetragonal phase transition at a temperature $T_s > T_c$, which involves strong softening of acoustic phonons associated with the $c_{11} - c_{12}$ elastic modulus. The relevance of these soft phonon modes for an increased $T_c$ were discussed early on [4]. On the other hand, the impact of 2$^{nd}$ order structural phase transitions on emerging superconductivity in charge-density-wave ordered materials has renewed the interest in the corresponding soft phonon modes [5-8].

Phonon measurements using inelastic neutron scattering (INS) have focused mainly on the high-$T_c$ materials Nb$_3$Sn and, to a lesser extent, on V$_3$Si. In the latter, the incoherent neutron scattering cross section of V strongly reduces the capability of INS to detect coherent one-phonon excitations. As already indicated above, strong phonon softening behavior was found generally in Nb$_3$Sn on cooling to the martensitic transition with the most pronounced effects in the transverse acoustic mode propagating along the [110] direction [9-11]. For Nb$_3$Sn, Axe and Shirane [12] reported first measurements on superconductivity-induced changes of the phonon line width when the phonon energy is smaller than the superconducting energy gap $2\Delta$. For V$_3$Si strong soft-mode behavior could be confirmed in the transverse acoustic (TA) mode propagating along the [110] direction and, to a lesser extent, also in other modes [13]. Yethiraj [14] reported more recently the appearance of a new peak in the superconducting phase similar to results for the rare-earth nickel-borocarbide superconductors but the data quality did not allow a detailed analysis.

In our work we overcome the above described problems by using high-energy-resolution inelastic x-ray (HERIX) scattering which shows strong coherent scattering intensities in V$_3$Si and requires only small sample volumes. We compare our results with detailed *ab-initio* lattice dynamical calculations based on density-functional-perturbation theory (DFPT) from which we derive parameters relevant for electron-phonon coupling (EPC) in V$_3$Si and estimate $T_c$. We demonstrate that phonon lineshapes obtained in the superconducting phase are asymmetric and well-explained by a theory proposed by Allen et al. [15].

## II. Theory

The calculations reported in this paper were performed in the cubic high-temperature phase of V$_3$Si in the framework of density functional theory with the mixed basis pseudopotential method [1]. Scalar-relativistic norm-conserving pseudopotentials of Vanderbilt-type were constructed for V$_3$Si [2], treating 3*s* and 3*p* semicore states of V explicitly as valence states. The mixed-basis scheme uses a combination of local functions and plane waves for the representation of the valence states [1], which allows for an efficient treatment of the fairly deep norm-conserving pseudopotentials. Local basis functions of *s*, *p*, and *d* type at V sites were supplemented by plane waves up to a kinetic energy of 25 Ry. For the exchange correlation functional, the generalized gradient approximation in the parametrization of Perdue-Burke-Ernzerhof (GGA-PBE) [3] was applied.

Phonon properties and EPC matrices were calculated using the linear response technique or density functional perturbation theory (DFPT) [4] in combination with the mixed-basis pseudopotential method [5]. Brillouin zone (BZ) integrations were performed by **k**-point sampling in conjunction with the standard smearing technique [6] employing a Gaussian broadening of 0.2 eV. Cubic 8x8x8 meshes corresponding to 20 **k**-points in the irreducible BZ (IBZ) were used for phonon calculations as well as for structural optimization, while a denser 16x16x16 mesh with 120 **k**-points in the IBZ was used for the computation of EPC matrix elements, which involve slowly converging Fermi-surface averages.

## III. Experiment

The IXS experiments were carried out at the 30-ID beamline, HERIX spectrometer, at the Advanced Photon Source, Argonne National Laboratory, with a focused beam size of $15 \times 32 \ \mu m^2$. The incident energy was 23.724 keV [16] and the horizontally scattered beam was analyzed by a set of spherically curved silicon analyzers (Reflection 12 12 12) [17]. The full width at half maximum (FWHM) of the energy and wave vector space resolution was about 1.5 meV and 0.066 Å$^{-1}$, respectively, where the former is experimentally determined by scanning the elastic line of a piece of Plexiglas and the latter is calculated from the experiment geometry and incident energy. The components ($Q_h$, $Q_k$, $Q_l$) of the scattering vector are expressed in reciprocal lattice units (r.l.u.) ($Q_h$, $Q_k$, $Q_l$) = (h*2$\pi$/a, k*2$\pi$/a, l*2$\pi$/a) with the lattice constants a = 4.72 Å of the cubic unit cell. Measurements were made in the constant-wave vector **Q** mode, *i.e.* as energy scans at constant wave vector **Q** = **τ** + **q**, where **τ** is a reciprocal lattice point and **q** the reduced wave vector. We used a high-quality single crystal sample grown at Oak Ridge National Laboratory [18], of about 50 mg (4 x 1 x 0.05 mm³) with transition temperatures for the structural and superconducting transition of $T_s =$ 18.9 K and $T_c = 16.8$ K determined by thermal expansion measurements [Fig. 2(a)]. The sample was mounted in a closed-cycle refrigerator and measurements reported here were done at various temperatures 4 K $\leq T \leq$ 300 K.

Measured energy spectra were fitted using a pseudo-Voigt function for the elastic line and a damped harmonic oscillator (DHO) function [19] for the phonon peaks. The fit function was convoluted with the fit of the experimental resolution function. The DHO function is described in equation 1:

$$S(Q,\omega) = \frac{[n(\omega)+1]Z(Q)4\omega\Gamma/\pi}{[\omega^2-\widetilde{\omega}_q^2]^2+4\omega^2\Gamma^2} \quad (1)$$

where $Q$ and $\omega$ are the wavevector and energy transfer, respectively, $n(\omega)$ is the Bose function, $\Gamma$ is the imaginary part of the phonon self-energy, $\widetilde{\omega}_q$ is the phonon energy renormalized by the real part of the phonon self-energy and $Z(Q)$ is the phonon structure factor. This function covers the energy loss and energy gain scattering by a single line shape. The intensity ratio of the phonon peaks at $E = \pm\omega_q$ is fixed by the principle of detailed balance. The energy $\omega_q$ of the damped phonons is obtained from the fit parameters of the DHO function by $\omega_q = \sqrt{\widetilde{\omega}_q^2 - \Gamma^2}$ [20].

## IV. Results – Theory

Figure 1 shows the calculated lattice dynamics. For clarity, we only plot phonon dispersions with particular polarizations (indicated in the panels) along four different high-symmetry directions. Furthermore, we plot the calculated electronic contributions to the phonon linewidth $\gamma$ as vertical bars for the respective phonon modes in Figs. 1(a)-(d). The generalized phonon density of states (PDOS) is shown [Fig. 1(e)] along with the Eliashberg function $\alpha^2F(\omega)$ and the total EPC constant $\lambda$ [Fig. 1(f)]. The phonon spectrum is clearly divided into a lower-energy part up to 33 meV including the steep acoustic phonons and a set of flat phonon dispersions just above 40 meV. The partial PDOSs [Fig. 1(e)] show that the high energy phonons are dominated by Si movements whereas the ones up to 33 meV are essentially pure V vibrations. The calculated PDOS is in good agreement with results from powder inelastic neutron scattering [21] [inset in Fig. 1(e)]. Only the high-energy phonons appear at slightly smaller energies in experiment. Using IXS we find similarly good agreement with data taken at room temperature [squares in Figs. 1(a)-(d)].

In our calculations, the V vibrations carry strong EPC evident from the large calculated phonon linewidths and the energy dependence of $\alpha^2F(\omega)$ [Fig. 1(e)]. In total, 85% of the contributions to $\lambda = 1.255$ originate from phonons at intermediate energies $16\text{ meV} \leq E \leq 33\text{ meV}$ [Fig. 1(f)]. With the calculated effective phonon energy $\omega_{eff} = 22.334$ meV, we solve the full gap equations and estimate the superconducting transition temperature $T_c$. Only an estimate of $T_c$ can be given, because the value of the effective electron-electron interaction potential $\mu^*$ is not calculated within DFPT. Using typical values of $\mu^* = 0.1 - 0.15$ we estimate $T_c = 20 - 26$ K where the higher $T_c$ corresponds to the lower value of $\mu^*$. Hence, an additional boost of EPC by the soft phonon mode of the structural phase transition on cooling towards $T_s$ is not needed to explain the superconducting transition temperature in V$_3$Si. In fact, the calculated contribution to $\lambda$ for phonon energies of $E \leq 10$ meV is less than 3%. Hence, even the strong increase of the linewidths of low-energy phonons which we report in the experimental section cannot result in a decisive role of the soft phonon mode for the high value of $T_c = 16.8$ K in V$_3$Si.

## V. Results – Experiment

Phonon spectroscopy has tremendously benefitted from the implementation of high energy-resolution inelastic x-ray scattering which enabled the measurements on small single crystals as well as on samples with large incoherent neutron cross section. Seminal materials with strong EPC could only recently be investigated due to their small sample sizes [22,23] including high-pressure experiments in diamond-anvil cells [6,7]. Here, we employ another advantage of IXS, the ability to detect phonons in materials featuring the element vanadium, which has a vanishing coherent neutron scattering cross section.

Before we present our results it is important to note that V$_3$Si features a high sensitivity of the structural and superconducting transition temperatures, $T_s$ and $T_c$, respectively, with respect to sample preparation. In many samples both $T_s$ and $T_c$ are observed always with $T_s > T_c$. But there are also samples featuring only a superconducting transition at $T_c$. Measurements of the sound velocity in samples with only $T_c$ indicate that $T_s$ would be lower than $T_c$ but the cubic structure is stabilized in the superconducting state [4]. In fact, specific heat data in a non-transforming sample of V$_3$Si revealed $T_{s,B>0} \approx 15$ K at finite magnetic fields smaller than the zero-field superconducting transition temperature $T_{c,B=0} = 16.2$ K [24].

In order to determine the transition temperatures in our sample we have performed dilatometry measurements and the obtained thermal expansion coefficient $\alpha(T)$ is plotted in Fig. 2(a). It shows a sharp spike at the structural phase transition temperature $T_s = 18.9$ K and another kink at the superconducting phase transition temperature at $T_c = 16.8$ K.

In the first subsection we report the evolution of various acoustic phonon modes on cooling towards the phase transition temperatures $T_s$ and $T_c$, i.e., down to 20 K. In the second part, we report on the changes of the phonon lineshape of some of these phonons on further cooling into the superconducting phase. We demonstrate that the observed changes can be very well described by a theory put forward by P.B. Allen and co-workers [15].

### V.1 Phonons in the normal state

We show IXS data taken at three different wavevector positions along the [100], [110] and [111] directions at $T = 20$ K $- 30$ K and room temperature in Figs. 2(b)-(d), 2(e)-(g) and 2(h)-(j), respectively. The data demonstrate a clear softening (reduction of the phonon energy) and broadening (increase of the linewidth $\Gamma$ of the intrinsic phonon lineshape approximated by a DHO function - after corrections for the instrumental resolution have been applied) of the respective longitudinal acoustic (LA) and TA modes on cooling. Note that phonon structure factors for both the LA and TA modes dispersing along the [110] are significant at $Q = (4-h, h, 0)$, Hence, both modes can be observed in close vicinity in Fig. 2(e). The results for the phonon energies and linewidths for the dominating modes in panels (b)-(j) are summarized in Figs. 2(k)-(m) and 2(n)-(p), respectively, compared to corresponding

DFPT calculations (lines). For the LA mode dispersing along the [100] direction, temperature-dependent softening [Fig. 2(k)] and broadening [Fig. 2(n)] is observed all the way to the Brillouin zone boundary. Interestingly, the results at room temperature for both phonon energies and linewidths are in good agreement with our calculations, whereas the low-temperature linewidths are about 1 meV broader.

We summarize analogous measurements of the in-plane polarized TA phonon branch along the [110] direction [Figs. 2(l)(o)] and the TA mode dispersing along the [111] direction [Figs. 2 (m)(p)]. While the latter shows (within the limited amount of data points) similar behavior as the LA mode, *i.e.*, overall softening and broadening, the TA mode in the [110] direction shows a clear soft mode behavior for a structural phase transition with ordering wavevector $q_s = 0$. The dispersion shows strong deviation from linear behavior in agreement with strongly decreasing elastic constants (which are proportional to the square of the slope of the soft mode dispersion at $q = 0$) reported by ultrasound measurements [4]. The calculated EPC of this mode is, in fact, smaller than that of the LA mode. The predicted values of $\gamma < 0.5$ meV are in reasonable agreement with our room temperature results for the half-widths at half maximum (HWHM). Yet, the increase of the experimental values of Γ on cooling to $T = 20$ K is even larger than for the LA mode with absolute values of up to 2.5 meV [Fig. 2(l)].

The huge increase of Γ of the TA mode is obvious from the IXS data taken at $Q = (3.85, 0.15, 0)$ and $T = 300$ K and 20 K [Fig. 3(a)]. The other visible phonon mode is the LA mode along the [110] direction, which has a non-vanishing structure factor. Interestingly, this mode shows neither softening nor broadening. We have performed a detailed temperature dependence of the TA soft phonon mode exhibiting strong phonon renormalization on cooling [Fig. 3(b)]. Note that we do not expect a full softening to zero energy at the investigated finite wavevector as the ordering wavevector is $q = 0$ for the martensitic phase transition in V$_3$Si. Still the damping ratio of the DHO function reaches up to $\Gamma/\widetilde{\omega}_q = 0.7$.

**V.2 Phonons in the superconducting state**

IXS measurements at $Q = (3.85, 0.15, 0)$ and T = 4 K reveal a sharp peak located near the peak position of the TA mode observed at T = 20 K [Fig. 3(a)]. We show results based on the same analysis with DHO functions as used for the normal state also for temperatures $T < T_c$ [Fig. 3(b)] although we will explain in the following that, in general, phonons with strong EPC and energies comparable to the value of the superconducting energy gap 2Δ acquire a more complex lineshape in the superconducting state. However, we want to connect to previous reports of phonons in A15 compounds and other conventional superconductors, in which authors have used this simpler approach. It has been reported that if the phonon energy is smaller than 2Δ, it loses its coupling (and intrinsic phonon linewidth Γ) and acquires a resolution limited lineshape, *e.g.*, in Nb$_3$Sn [12], elemental Nb [25] and Pb [26,27] using inelastic neutron scattering.

Results from our analysis for V$_3$Si shown in Fig. 3(b) reveal a similar narrowing of the approximated DHO function on cooling below $T_c$ by a factor of ten whereas the peak position stays roughly constant. The energy of the TA mode is around $3 − 3.5$ meV for $T \leq 20$ K and, hence, arguably well-below the reported superconducting gap value of $2\Delta = 3.5\ k_B T_c \approx 5$ meV [28,29]. However, the TA mode measured just above $T_c$ exhibits such a large linewidth Γ that a sizeable part of the phonon intensity is located above 5 meV [blue circles in Fig. 3(a)].

Indeed, data taken in the superconducting phase at $T = 4$ K and $Q = (3.85, 0.15, 0)$ are incompatible with this analysis [Fig. 4(a)]: apart from the narrow peak at 3 meV and 13 meV we observe a broad intensity distribution for energy transfers from 5 meV all the way up to 10 meV. While this intensity is not peak-shaped it also cannot be explained as tails of the two narrow peaks due to the finite experimental energy resolution.

In the following, we will demonstrate that a theory provided by Allen et al. [15] semi-quantitatively predicts the phonon lineshapes in the superconducting phase [green solid line in Fig. 4(a)] and reveals clear anisotropies of the value of 2Δ for electronic states connected by phonon wavevectors with different directions in reciprocal space. In fact, we have used this theory to explain inelastic neutron scattering results in YNi$_2$B$_2$C [30,31] (which originally motivated the theoretical work) as well as in elemental Nb [32].

The explicit form for the phonon lineshape in a superconductor is given by [15]

$$S(Q,\omega) = \frac{4\omega r_Q \omega_Q^2 \gamma_s/\gamma_Q}{\left[\omega^2 - \omega_Q^2 - 2\omega_Q^2 r_Q \mathrm{Re}(\delta\Pi)/\gamma_Q\right]^2 + \left[2\omega r_Q \omega_Q \gamma_s/\gamma_Q\right]^2}$$

where $\gamma_s$ is the imaginary part of the phonon self-energy $\Pi_s$ in the superconducting state, $\omega_Q$ and $\gamma_Q$ are the phonon energy $E_{phon}$ and linewidth $\Gamma_{phon}$ in the normal state defining the dimensionless ratio $r_Q = \gamma_Q/\omega_Q$. The calculation depends explicitly on three parameters: (1) The phonon energy $E_{phon} \equiv \omega_Q$, (2) the phonon linewidth $\Gamma_{phon} \equiv \gamma_Q$ (both determined at $T \gtrsim T_c$), and (3) the value of the superconducting gap 2Δ at the respective temperature of the measurement $T < T_c$. Two of the parameters, $E_{phon}$ and $\Gamma_{phon}$, are determined by the DHO fit to the experimental data at T = 20 K shown in Fig. 4(a). Explicitly, we use $E_{phon} = 3.0$ meV and $\Gamma_{phon} = 2.7$ meV [see Fig. 3(b)] for our calculation for $Q = (3.85, 0.15, 0)$. Hence, 2Δ is the only free parameter in this analysis.

In Figure 4(a) we show the fit (blue solid line) of the experimental data taken in the normal state at $T = 20$ K and $Q = (3.85, 0.15, 0)$. The blue dashed line with a maximum close to 3 meV denotes the DHO fit of the TA mode determining the values of $E_{TA} = (3.0 \pm 0.3)$ meV and $\Gamma_{TA} = (2.7 \pm 0.2)$ meV entering the calculations. The green solid line represents the prediction based on [15] for the phonon lineshape in the superconducting state with $2\Delta = 5.2$ meV exhibiting reasonable agreement with the experimental data at $T = 4$ K. In the following, we will explain in detail how we deduce the low-temperature phonon lineshape.

Results of the model calculation using the above given values of $E_{TA} = 3.0$ meV and $\Gamma_{TA} = 2.7$ meV and $2\Delta = 5.2$ meV are shown in the inset of Fig. 4(b). The prediction is a very sharp resonance-like peak well below $2\Delta$ and a broad intensity distribution for $E \geq 2\Delta$ [red dashed line in inset of Fig. 4(b)]. In order to compare with our experimental data we convolute the calculation with the experimental resolution represented by a pseudo-voigt profile with a linewidth of 1.5 meV (FWHM). The resulting lineshape [black dash-dotted line in inset of Fig. 4(b)] shows already a close resemblance to the experimental data at $T = 4$ K in the energy range $2$ meV $\leq E \leq 9$ meV. The main panel of Fig. 4(b) shows the variation of the calculated phonon lineshape on increasing $2\Delta$ from 5.2 meV to 6.0 meV (black dash-dotted and red short-dashed lines, respectively). Additionally we plot the DHO function fitted to the TA mode at $T = 20$ K (upper blue dashed blue line) and the same DHO function but with a temperature factor equivalent to $T = 4$ K (lower blue dashed blue line) simulating the cooling-induced changes of the phonon intensity as if the phonon energy and linewidth remained constant. From this, it is clear that the experimentally observed changes [Fig. 4(a)] are a superposition of normal temperature- and superconductivity-induced effects. The former are visible as the difference between the DHO functions [blue dashed blue lines in Fig. 4(b)] whereas the latter is represented by the changes between the low-temperature DHO function (blue dashed blue line) and the calculated phonon profiles for $2\Delta = 5.2$ meV and 6.0 meV (black dash-dotted and red short-dashed lines, respectively) in Fig. 4(b). In order to compare our results with the raw data shown in Figure 4(a), we plot the sum of the calculated curve for $2\Delta = 5.2$ meV and the fit functions for the optic mode at 13 meV, the elastic line and the constant background obtained at $T = 20$ K as green solid line in Figure 4(a). We find a remarkable agreement with the experimental data considering that there is only one free parameter in the calculation, $2\Delta$. The agreement is not quantitative as the sharp peak in the calculation is located at a slightly smaller energy transfer and has a larger intensity compared to the experimental data.

In order to discuss the differences between experiment and calculation in more detail we plot the difference in the count rate between data taken at $T = 20$ K and 4 K in Fig. 4(c) in the energy range $0 \leq E \leq 10$ meV. The black dash-dotted and red short-dashed lines denote the difference between the DHO function approximated for the TA mode at $T = 20$ K and the calculated spectra for $2\Delta = 5.2$ meV and 6.0 meV, respectively, shown in Fig. 4(b). The above mentioned small disagreement between the peak positions in experiment and calculation is visible in the shifted minima of the difference curves. Increasing the value of $2\Delta$ from 5.2 meV to 6.0 meV shifts the calculated minimum toward the position of the observed one. However, the discrepancy between the absolute values at the position of the minima becomes larger. Hence, larger values of the superconducting energy gaps cannot remedy the quantitative disagreement.

Figures 5(a)-(h) show data for four additional wave vectors along the [110], [100] and [111] directions in $\mathbf{Q}$ along with the corresponding experimental and calculated difference curves [Figs. 5(i)-(l)]. Again, we see that the calculations describe the observed effects qualitatively but have discrepancies in the position and/or the strength of the minima in the difference plots. Here, it is instructive to discuss different scenarios in the model of Allen et al. [15] with regard to the values of the phonon energy in the normal state $E_{NS}$ and the superconducting energy gap $2\Delta$ relative to each other.

(1) $E_{NS} < 2\Delta$:
Nearly all of the phonon spectral weight is below $2\Delta$. Hence, the phonon loses its coupling because it cannot excite electrons from below to above the superconducting energy gap. The peak energy in the superconducting phase is practically the same as above $T_c$. In IXS we find a resolution limited peak at $E_{NS}$. Examples for this scenario are the data taken at $\mathbf{Q} = (3.9, 0.1, 0)$ with $E_{NS}/2\Delta \approx 1/4$ [Fig. 5(a)(e)]. While the same phonon mode at $\mathbf{Q} = (3.85, 0.15, 0)$ exhibits a phonon energy $E_{NS} \approx 0.5 \times 2\Delta$, the very large linewidth results in a sizeable intensity also above $2\Delta$. Hence, the lineshape is a mix between scenarios (1) and (2), where the peak position at $T = 4$ K is that of the normal-state phonon [scenario (1)] but a broad tail above $2\Delta$ remains [scenario (2)].

(2) $E_{NS} \approx 2\Delta$:
The broadened phonon has large parts of its spectral weight at energies below and above $2\Delta$. Whereas the spectral weight below $2\Delta$ condenses into a resonance-like peak dominating the low-temperature spectrum, the part at energies above $2\Delta$ keeps the broad distribution over energy because the energy is still sufficient to excite electrons and, hence, electron-phonon coupling effects remain active even below $T_c$. Experimentally, we observe an energy-shift of the peak intensities on cooling below $T_c$ but a broad intensity distribution remains at energies above $2\Delta$. Examples for this scenario are the data taken at $\mathbf{Q} = (4.1, 0, 0)$ [Fig. 5(b)(f)] and $(2.1, 2.1, -1.9)$ [Fig. 5(c)(g)] with $E_{NS}/2\Delta \approx 1$.

(3) $E_{NS} > 2\Delta$:
Only a minor part of the phonon spectral weight is located at energies smaller than $2\Delta$ and is shifted towards energies $E \geq 2\Delta$. Depending on the specific strength of the effect, IXS reveals either a shoulder on the low energy side of the normal-state phonon or, if the effect is stronger, a peak at lower energies. However, the sharp feature near $2\Delta$ and the broad intensity distribution at the normal state phonon energy are of similar size. This scenario is nicely illustrated by measurements at $\mathbf{Q} = (2.15, 2.15, -1.85)$ [Fig. 5(d)(h)].

Figures 5(i)-(l) show the observed and calculated differences, the latter computed for $2\Delta = 4.2$ meV (red dash-dotted lines) and 5.6 meV (orange dashed lines). For data taken at $\mathbf{Q} = (3.9, 0.1, 0)$ scenario (1) applies and, hence, the peak position in the superconducting phase is not very sensitive to the value of $2\Delta$ [Fig. 5(i)]. The position in energy of the minimum of the difference curves varies only by 0.09 meV when $2\Delta$ changes by 1.4 meV. Yet, the calculations do not agree quantitatively with the observed minimum, which is shifted to slightly higher energies. We believe that this is a signature of the soft mode character of the TA mode related to the structural phase transition. In spite of superconductivity, the soft mode should harden on cooling below $T_s = 18.9$ K when the tetragonal structure is stabilized. Hence, the analysis shown in Fig. 5(i) is not suited to determine the value of $2\Delta$. The same mechanism might be the origin for the similarly observed shift in the analysis for $\boldsymbol{Q} = (3.85, 0.15, 0)$ [Fig. 4(c)].

Data shown in Figs. 5(j)-(l) are better suited because the respective modes are not as sensitive to the structural phase transition and the peak positions in the spectra taken at $T = 4$ K have a closer link to $2\Delta$ according to scenarios (2) and (3). Correspondingly, the shift between the calculated minima is a factor 4-5 larger for these wavevectors. From our analysis, we find that minima positions are best described using $2\Delta = 5.6$ meV at $\boldsymbol{Q} = (4.1, 0, 0)$ [Fig. 5(f)] but $2\Delta = 4.2$ meV at wavevectors along the [111] direction [Figs. 5(g)(h)]. Hence, the different parts of the Fermi surface sampled by phonons modes along the [100] and [111] directions show an anisotropic energy gap $2\Delta(\boldsymbol{q})$.

## VI. Discussion

The interplay of the martensitic and superconducting transitions in A15 compounds was already discussed in the 1970s [4]. However, it has become evident in the last decade that several layered materials featuring a structural phase transition connected to a charge-density wave (CDW) host superconducting phases if the structural phase transition is suppressed, e.g., by pressure [33-38] or intercalation [39,40]. Here, we want to point out similarities between our results for V$_3$Si and observations on the lattice dynamical properties of the seminal layered compound *2H*-NbSe$_2$ featuring CDW order ($T_{CDW} = 33$ K) [41] and superconductivity ($T_c = 7.2$ K) [42] at ambient pressure.

*2H*-NbSe$_2$ exhibits strong coupling to the electronic excitations for a large number of phonon branches throughout reciprocal space [43]. Based on *ab-initio* lattice dynamical calculations [7], it was concluded that the soft mode of the CDW order, *i.e.*, a LA phonon at $\boldsymbol{q}_{CDW} = (0.329, 0, 0)$ contributes little to the overall EPC constant $\lambda$. Furthermore, the CDW gap opens only on a small part of the Fermi surface [44,45] explaining the metallic character of *2H*-NbSe$_2$ in the CDW ordered phase at $T_c < T < T_{CDW}$ [42].

Hence, the properties of the CDW soft phonon mode do not have a large impact on superconductivity – similar to our result for V$_3$Si. We note here that we did not investigate phonons with predicted strong EPC in the energy range of 20 meV $\leq E_{phon} \leq$ 30 meV (see Fig. 1). However, the strong EPC of a large number of phonons in this energy range was observed by phonon density of states measurements in V$_3$Si powder samples: Delaire, Lucas, Muñoz, Kresch and Fultz [21] reported an anomalous hardening of a phonon band in the range of 26 meV $\leq E \leq$ 28 meV on heating from $T = 10$ K up to $T \approx 600$ K. On further heating this band showed the usually expected softening because of thermal expansion. The hardening was interpreted as a sign of strong EPC [21].

It is interesting to discuss the implications of the lattice dynamics on the phase competition in V$_3$Si, *e.g.*, as function of pressure. For V$_3$Si samples with $T_s > T_c$ ($p = 0$), $T_s$ decreases and $T_c$ increases with pressure [46,47]. Yet, it is an open question, whether the observed broad maximum of $T_c$ centered at $p \approx 8$ GPa [47] corresponds to the critical pressure of the martensitic phase transition line in the hypothetical absence of superconductivity. The above given discussion on the small impact of the soft phonon mode on the superconducting properties of V$_3$Si suggests that the evolution of $T_c$ might be essentially independent of that of the martensitic transition. However, $T_s$ is suppressed under pressure as soon as it reaches the respective value of $T_c$ since the opening of the superconducting energy gap suppresses all EPC in the soft phonon mode which drives the martensitic phase transition [see Fig. 3(b)].

Last, we want to discuss the observation of the superconductivity-induced changes of the phonon lineshape. While similar observations have been made with neutron scattering in YNi$_2$B$_2$C [30,31] and elemental Nb [32], our results are the first clear evidence for such a behavior from inelastic x-ray scattering. On the other hand, the report from Axe and Shirane [12] on the "Influence of the superconducting energy gap on the phonon linewidth in Nb$_3$Sn" – another A15 compound - is, to our knowledge, the very first report on this topic albeit with neutrons. Hence, we extracted the data showing the transverse acoustic phonon in Nb$_3$Sn at $\boldsymbol{q} = (0.18, 0.18, 0)$ from Fig. 1 in [12]. Figure 6 shows our analysis using the same procedure as presented above for V$_3$Si. While the effect seems to be overestimated, the model theory can explain the remaining spectral weight at energies slightly above $2\Delta = 4.3$ meV. It also explains very well that the peak in the superconducting phase occurs below the normal-state phonon energy.

## VII. Conclusion

In summary, we have made an inelastic x-ray scattering investigation of the lattice dynamics of V$_3$Si. Comparison of the experimental results to *ab-initio* lattice dynamical calculations suggest only a small impact of the soft phonon mode of the martensitic transition on the superconducting properties. Measurements of low-energy acoustic phonon modes on cooling into the superconducting phase display pronounced changes of the spectral weight distribution. We show that these changes are signatures of the opening of the superconducting energy gap, which can be explained by a model theory put forward by Allen, Kostur, Takesue and Shirane [15]. Analysis of a long-time published data set showing $T_c$- induced effects for phonons in the A15 compound $Nb_3Sn$ demonstrates the general applicability of this model.

**VIII. Acknowledgements**

D. Z. and F.W. were supported by the Helmholtz Association under contract VH-NG-840. This research used resources of the Advanced Photon Source, a U.S. Department of Energy (DOE) Office of Science User Facility operated for the DOE Office of Science by Argonne National Laboratory under Contract No. DE-AC02-06CH11357.

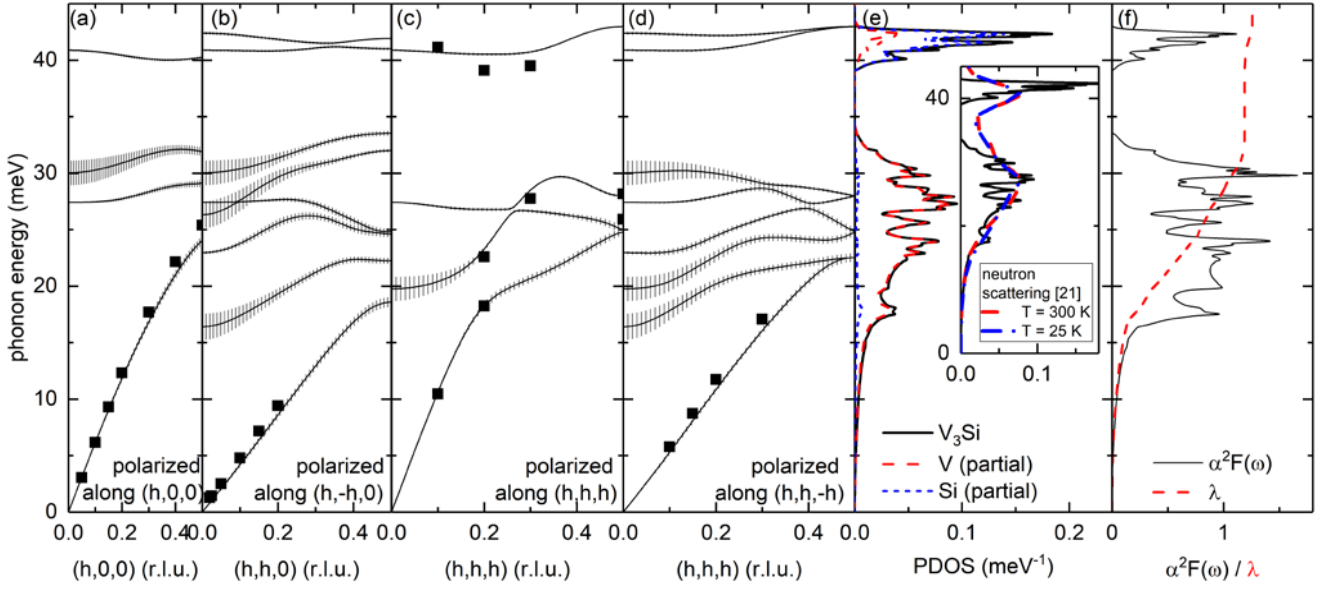

**FIG. 1:** First-principles lattice dynamical calculations of $V_3Si$ compared to the experimental results obtained at room temperature. (a)-(d) Calculated phonon dispersion lines (lines) compared to the observed phonon energies at room temperatures (squares). Vertical lines represent the calculated electronic contributions to the phonon linewidth of the respective mode. (e) Calculated total (solid line) and partial (dashed/dotted lines) phonon density of states. The inset shows a comparison to the results from inelastic neutron scattering for $T = 25$ K (blue dash-dotted line) and 300 K (red dashed line) published in Ref. [21]. (f) Calculated Eliashberg function $\alpha^2F(\omega)$ (black solid line) and electron-phonon coupling constant $\lambda$ (red dashed line).

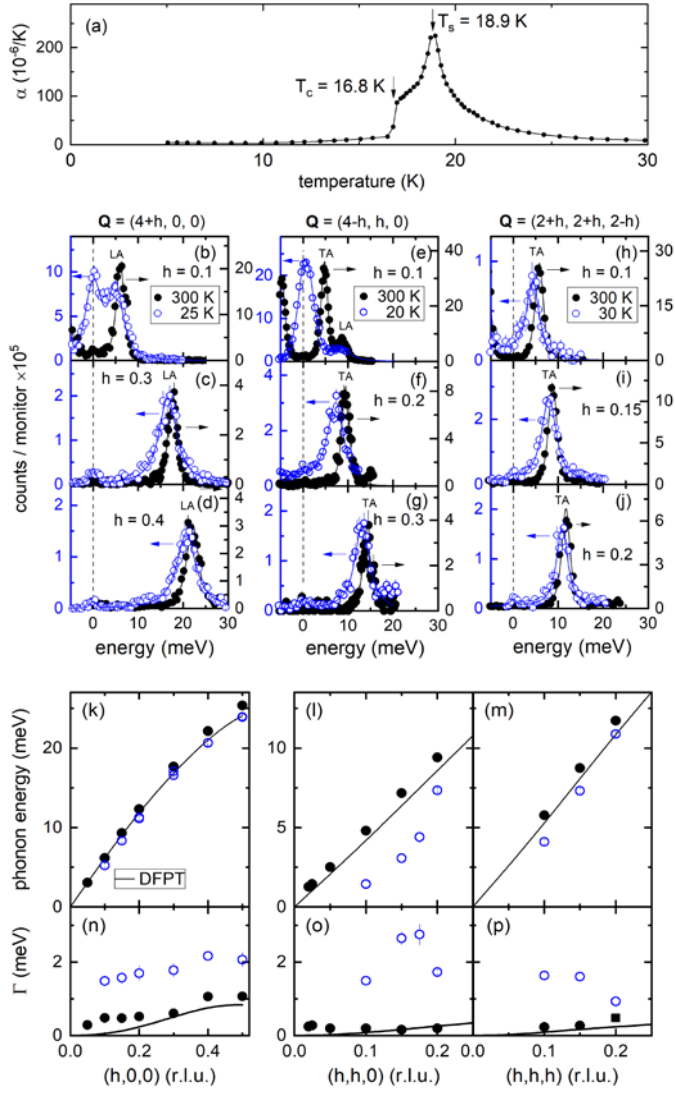

**FIG. 2.** (a) Temperature dependence of the thermal expansion coefficient $\alpha = \frac{1}{L} \cdot \frac{\Delta L}{\Delta T}$ derived from dilatometry measurements of the same single crystal used for the inelastic x-ray scattering experiments. The sharp peak and the lower-temperature kink indicate the structural and superconducting phase transition in our sample at $T_s = 18.9$ K and $T_c = 16.8$ K, respectively. (b)-(j) Evolution of IXS raw data (normalized to monitor) on cooling from room temperature (black dots, right-hand scale) to low temperatures $T = 20$ K -30 K (blue circles, left-hand scale). We observe an LA phonon dispersing along the [100] direction (b)-(d) and a TA phonon dispersing along the [111] direction (h)-(j). The focus along the [110] direction (e)-(g) was on the corresponding TA phonon but the LA mode has also a finite structure factor [see (e)]. Lines are fits to the data using a DHO function for phonon peaks convoluted with the experimental resolution and a pseudo-Voigt function for the elastic line. (k)-(m) Phonon energies and (n)-(p) linewidths deduced from the above presented raw data (b)-(j) (same color/symbol code) compared to the corresponding *ab-initio* calculations (lines).

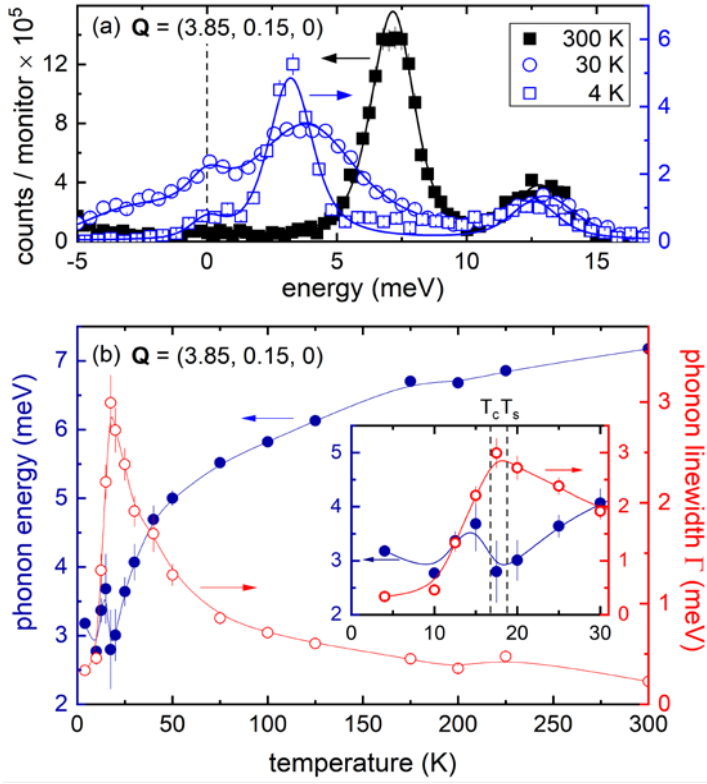

**FIG. 3.** (a) Raw data of IXS measurements taken at T = 300 K (black dots, right-hand scale), 30 K (blue circles, left-hand scale) and 4 K (blue squares, left-hand scale) for the wavevector **Q** = (3.85,0.15,0). Lines are fits to the data using a DHO function for phonon peaks convoluted with the experimental resolution and a pseudo-Voigt function for the elastic line. (b) Temperature dependence of the phonon energy (blue dots) and linewidth (red circles) of the soft phonon mode at **Q** = (3.85,0.15,0) when analyzed with a DHO function at all temperatures (see text). The inset shows a zoom of the low-temperature range $T \leq 30$ K.

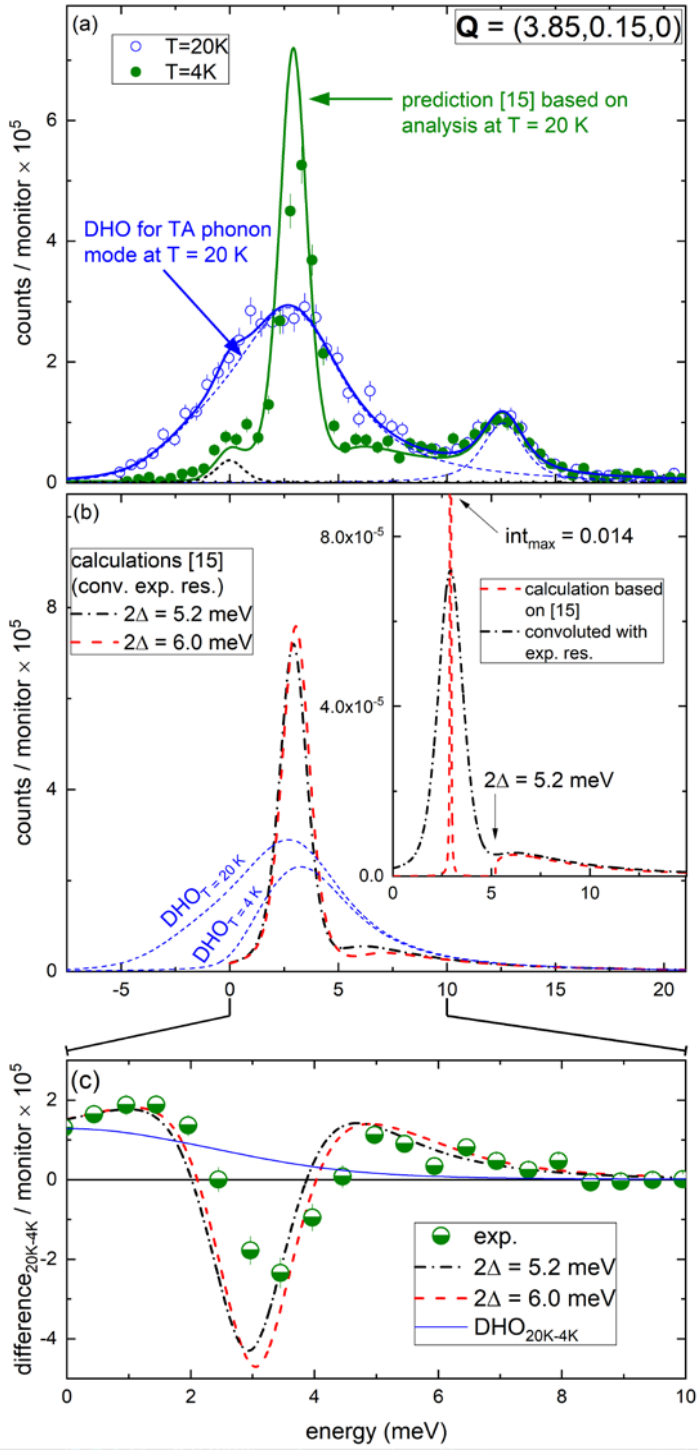

**FIG. 4.** Analysis of the superconductivity-induced changes of the phonon lineshape for the structural soft phonon mode observed at $Q = (3.85,0.15,0)$. (a) IXS raw data taken at $Q = (3.85,0.15,0)$ at $T = 20$ K (green dots) and 4 K (blue circles), *i.e.*, just above $T_s$ (= 18.9 K) and $T_c$ (= 16.8 K) and well-below $T_c$. The (blue) solid line denotes the fit of the data taken in the normal state consisting of two DHO functions for the TA and TO modes (blue dashed lines) along with a pseudo-Voigt function for the elastic line (black short-dashed line) and a constant experimental background (black dotted line). The green solid line represents a calculation of the phonon lineshape of the TA mode based on the theory of Allen et al. [15] to which the DHO function corresponding to the TO mode, the elastic line and the experimental background were added (see text). (b) Upper and lower blue dashed lines represent the DHO fit of the TA mode at $T = 20$ K and the same DHO function except that the temperature factor was set to the equivalent of $T = 4$ K, respectively. The black dash-dotted and red dotted lines are calculations based on the theory of Allen et al. [15] and on the fitting parameters obtained at $T = 20$ K for the TA mode [see (a)] for two different values of the superconducting energy gap $2\Delta$. The inset shows the calculation for $2\Delta = 5.2$ meV with (black dash-dotted line) and without (red dashed line) a convolution with the experimental resolution. (c) Difference between the raw IXS data shown in panel (a) (green half-dots) compared to the differences obtained by subtracting the calculated lines in panel (b) from the DHO fit for $T = 20$ K for the respective values of $2\Delta$ (dotted/dash-dotted lines). The blue solid line represents the difference between the DHO fit at $T = 20$ K and the curve obtained by setting the temperature factor of the DHO fit to the equivalent of $T = 4$ K.

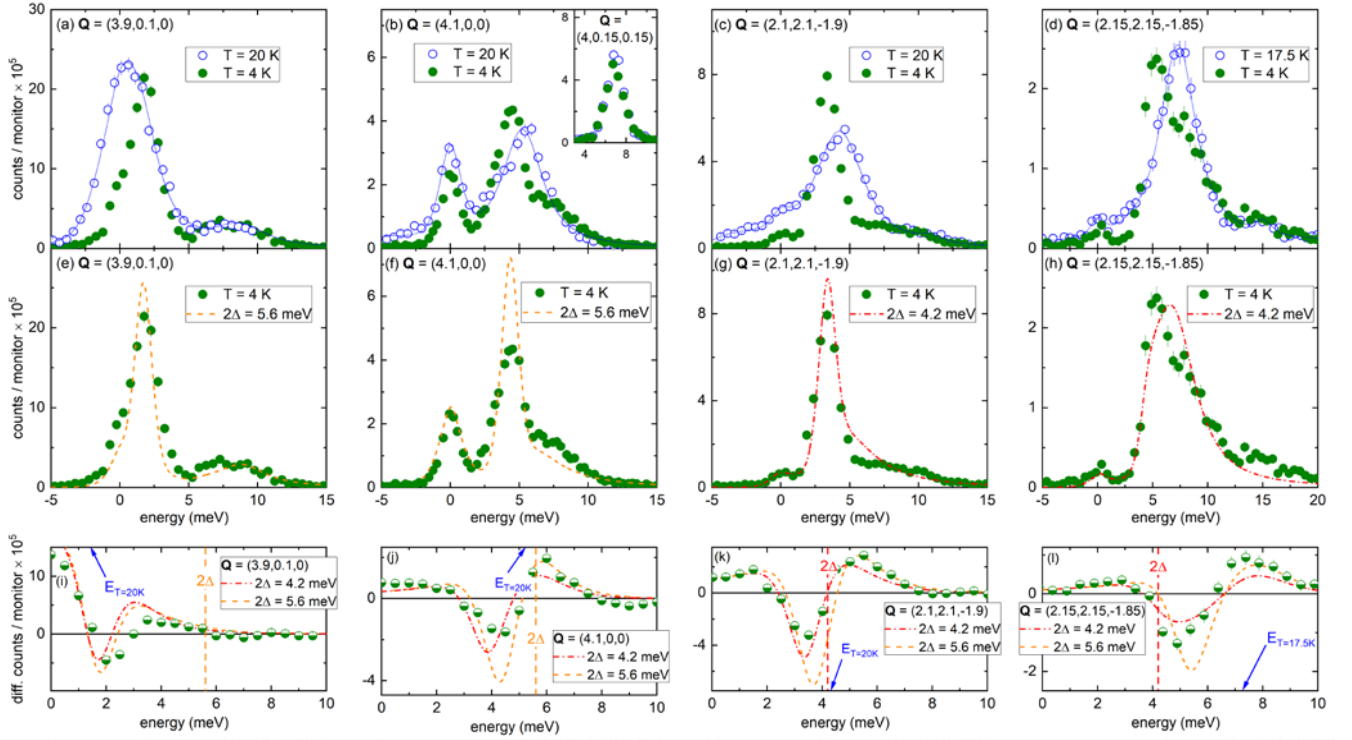

**FIG. 5.** (a)-(d) IXS raw data taken at four different wave vectors for $T = 20$ K (green dots) and 4 K (blue circles), *i.e.*, just above $T_s$ (= 18.9 K) and $T_c$ (= 16.8 K) and well-below $T_c$. Lines are fits to the normal state data including DHO functions for phonons (convoluted with the experimental resolution) and pseudo-voigt function for the respective elastic line. The inset in (b) shows data for $Q = (4,0.15,0.15)$ as an example with no superconductivity-induced changes of the phonon lineshape. (e)-(h) Comparison of the IXS data taken at T = 4 K (green dots) with the calculated phonon lineshape in the superconducting phase employing $2\Delta = 5.6$ meV (orange dashed lines) and 4.2 meV (red dash-dotted lines). (i)-(l) Differences between raw IXS data on cooling from T = 20 K to 4 K (half-green dots) compared to the respective theoretical predictions for $2\Delta = 4.2$ meV (red dash-dotted lines) and 5.6 meV (orange dashed lines) [see text and Fig. 4(c)] with the corresponding raw data shown in the top row of panels. Dashed vertical lines denote the color-coded value of $2\Delta$, for which calculations show the best agreement with experimental results. Blue arrows mark the observed phonon energies at the respective wave vectors at $T = 20$ K.

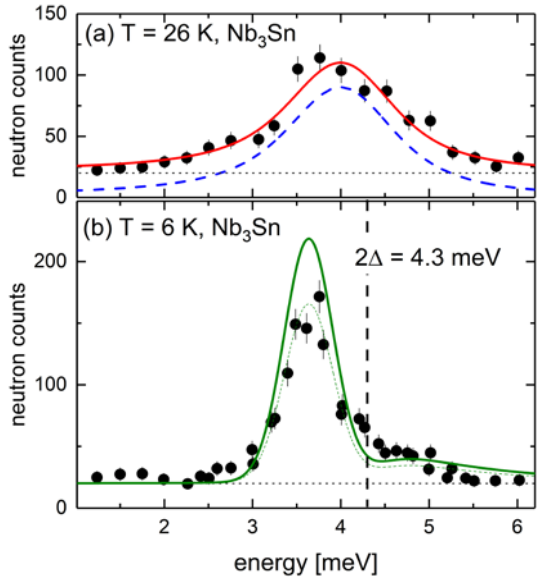

**FIG. 6.** Analysis of inelastic neutron scattering (INS) data taken for Nb$_3$Sn [12]. (a) Fit (red solid line) of a DHO function (blue dashed line) and an estimated experimental background (black dotted line) to INS data (black dots) taken above the superconducting transition temperature. (b) INS data taken in the superconducting phase (black dots) along with the calculated phonon lineshape (green solid line) using the model theory [15] on top of the experimental background (black dotted line). The green dotted line denotes the calculated result scaled by a factor of 0.73 on top of the background.